\documentclass[11pt,a4paper]{article}

\pdfoutput=1
\usepackage{jcappub}
\usepackage[latin1]{inputenc}
\usepackage{graphicx}
\usepackage{epsfig}
\usepackage{subfigure}
\usepackage{color}
\usepackage{comment}

\def\beq{\begin{equation}}
\def\eeq{\end{equation}}
\def\baq{\begin{eqnarray}}
\def\eaq{\end{eqnarray}}

\newcommand{\be}{\begin{equation}} 
\newcommand{\ee}{\end{equation}}
\newcommand{\bea}{\begin{eqnarray}} 
\newcommand{\eea}{\end{eqnarray}}
\newcommand{\nn}{\nonumber}
\newcommand{\bmp}{\noindent\begin{minipage}{16cm}}
\newcommand{\emp}{\end{minipage}\vskip 7mm} 
\def\lsim{\mathrel{\raise.3ex\hbox{$<$\kern-.75em\lower1ex\hbox{$\sim$}}}}
\def\gsim{\mathrel{\raise.3ex\hbox{$>$\kern-.75em\lower1ex\hbox{$\sim$}}}}

\newcommand{\intron}[1]{}

\subheader{\small{Preprint: HIP-2015-6/TH}}
\title{Inflationary Imprints on Dark Matter}

\author[a,b]{Sami Nurmi,}
\author[a]{Tommi Tenkanen}
\author[a]{and Kimmo Tuominen}

\affiliation[a]{University of Helsinki and Helsinki Institute of Physics, \\
                      P.O.~Box 64, FI-00014, University of Helsinki, Finland}
\affiliation[b]{Department of Physics, University of Jyv\"{a}skyl\"{a}, \\
P.O.~Box 35 (YFL), FI-40014 University of Jyv\"{a}skyl\"{a}, Finland} 
\emailAdd{sami.nurmi@helsinki.fi}
\emailAdd{tommi.tenkanen@helsinki.fi}
\emailAdd{kimmo.i.tuominen@helsinki.fi}

\abstract{We show that dark matter abundance and the inflationary
scale $H$ could be intimately related. Standard Model extensions
with Higgs mediated couplings to new physics typically contain extra
scalars displaced from vacuum during inflation. If their coupling to
Standard Model is weak, they will not thermalize and may easily constitute too
much dark matter reminiscent to the moduli problem. As an example we
consider Standard Model extended by a $Z_2$ symmetric singlet $s$
coupled to the Standard Model Higgs $\Phi$ via $\lambda \Phi^{\dag}\Phi s^2$. Dark matter
relic density is generated non-thermally for $\lambda \lesssim
10^{-7}$. We show that the dark matter yield crucially depends on
the inflationary scale. For $H\sim 10^{10}$ GeV we find that the
singlet self-coupling and mass should lie in the regime
$\lambda_{\rm s}\gtrsim 10^{-9}$ and $m_{\rm s}\lesssim 50$ GeV to
avoid dark matter overproduction.}


\begin{document}
\maketitle

%
\section{Introduction}

For the measured Higgs mass $m_{\rm h}\simeq 125$ GeV the Standard Model (SM) vacuum is metastable up to remarkably high energies $\mu_{\rm c} \sim 10^{11}$ GeV with a lifetime much longer than the age of the universe \cite{Degrassi:2012ry,Buttazzo:2013uya,Ellis:2009tp,Antipin:2013sga}. In the early universe the stability is crucially affected also by the large spacetime curvature. Radiative corrections to the Higgs potential necessarily induce the non-minimal curvature coupling $\xi\Phi^{\dag}\Phi R$ which during inflation when $R\sim H^2$ may easily dominate the potential \cite{Herranen:2014cua, Espinosa:2007qp,Herranen:2015ima}. Indeed, it has been shown that if $\xi(\mu_{\rm EW}) \gtrsim 0.1$
\cite{Herranen:2014cua} at the electroweak scale the SM Higgs
remains stable up to the highest
inflationary scale $H_{*}\sim 10^{14}$ GeV consistent with the tensor bound \cite{Ade:2015tva, Ade:2015xua}
without any need to include new physics.

New physics beyond SM is however implied by several cosmological
observations, such as inflation itself, dark matter and baryon
asymmetry. If the Higgs potential is not drastically
modified by new physics the Higgs generically is a light and
energetically subdominant field during inflation
\cite{Espinosa:2007qp, DeSimone:2012qr, Enqvist:2013kaa, Enqvist:2014bua, Herranen:2014cua, Fairbairn:2014zia, Kobakhidze:2013tn, Hook:2014uia, Espinosa:2015qea, Kearney:2015vba}.
(The attracting possibility of Higgs-driven inflation requires a
non-trivial deviation from the SM potential at high energies, such
as a large coupling to spacetime curvature
\cite{Bezrukov:2007ep,Bezrukov:2009db,Bezrukov:2010jz, Bezrukov:2014ipa, Bezrukov:2014bra, Hamada:2014iga}.)
Inflationary fluctuations displace the light Higgs from its vacuum
generating a primordial Higgs condensate \cite{Enqvist:2013kaa, DeSimone:2012qr}. The
resulting out-of-equilibrium initial conditions $h_*\sim H_{*}$ for
the hot big bang epoch may have significant observational
ramifications. Particle production from the time-dependent Higgs
condensate may for example generate baryon asymmetry
\cite{Kusenko:2014lra} or produce non-thermal dark matter
\cite{Enqvist:2014zqa}. A careful investigation of the observational
effects is of key importance and could reveal powerful new tests of
specific SM extensions at very high energy scales.

In this work we will investigate how the initial conditions set by
inflation affect the generation of dark matter abundance in the interesting
class of portal scenarios \cite{McDonald:1993ex,McDonald:2001vt,
Alanne:2014bra} where the SM fields feel new physics only through
Higgs-mediated couplings. As a representative example we will
consider a $Z_2$ symmetric scalar singlet $s$ coupled to the Higgs field by
$V_{\rm int}=\lambda_{\rm sh}^2 \Phi^{\dag}\Phi s^2$. This
simplified example captures some interesting features of the portal
models. The singlet constitutes a dark matter candidate
\cite{McDonald:1993ex,McDonald:2001vt,Burgess:2000yq,Cline:2013gha} and its
dynamics at electroweak phase transition could allow for
baryogenesis \cite{Espinosa:2011ax,Profumo:2007wc,Cline:2012hg}.
Analogously to Higgs also the singlet is generically a light field
during inflation and gets displaced from vacuum
\cite{Enqvist:2014zqa}. In this simple setup the out-of-equilibrium
initial conditions do not typically affect the electroweak
baryogenesis as the scalar condensates have either decayed or
diluted away by that time \cite{Enqvist:2014zqa}. They could however
significantly affect the dark matter abundance if the portal
coupling is very weak $\lambda_{\rm sh}\lesssim 10^{-7}$. In this
case the singlet never thermalizes with the SM fields. The singlet
dark matter originates entirely from non-thermal production of
singlet particles through the so called freeze-in mechanism
\cite{McDonald:2001vt, Hall:2009bx, Yaguna:2011qn, Merle:2013wta,
Klasen:2013ypa, Blennow:2013jba, Adulpravitchai:2014xna,
Merle:2014xpa, Elahi:2014fsa, Merle:2015oja, Kang:2015aqa}. As we will show in this paper, the presence of
scalar condensates significantly alters the previous estimates for
the efficiency of the process leading to novel interplay between
dark matter properties and initial conditions sensitive to the
inflationary scale. Moreover, the inflationary displacement of the
singlet may also lead to overproduction of dark matter which places
stringent constraints on viable singlet mass scales and
the values of its self-coupling.

The paper is organized as follows: In Section \ref{sec:model} we
define the $Z_2$ symmetric model and discuss the freeze-in mechanism
of dark matter production. In Section \ref{sec:condensates} we
discuss initial conditions set by inflation and the related moduli problem.
In Section \ref{sec:mechanism} we investigate the production of
singlet dark matter through the freeze-in accounting for the
presence of primordial scalar condensates in the Boltzmann
equations. We then present our main results for the dark matter
abundance and its dependence on scalar couplings and the
inflationary scale in Section \ref{totalYield}. Finally we summarize and discuss the results in Section
\ref{sec:conclusions}.

\section{Singlet dark matter and freeze-in without primordial condensates}
\label{sec:model}
We consider the SM extended to include a $Z_2$ symmetric
scalar singlet \cite{McDonald:1993ex,Burgess:2000yq,Cline:2013gha}
\be
V(\Phi,S)=m_{\rm h}^2\Phi^\dagger\Phi+\lambda_{\rm h}(\Phi^\dagger\Phi)^2+\frac{1}{2}m_{\rm s}^2
s^2+\frac{\lambda_{\rm s}}{4}s^4+\frac{\lambda_{\rm sh}}{2}(\Phi^\dagger\Phi)s^2\
.
\label{scalarpot}
\ee Here $\Phi$ is the Standard Model Higgs doublet and $s$ is a
real scalar singlet assumed to possess a $Z_2$ symmetry in order to
make it stable. Therefore, the singlet constitutes a dark matter
candidate. At low temperatures and in the unitary gauge, $\Phi =
\left(0,(\nu + h)/\sqrt{2}\right)^T$. We will assume that $m_{\rm s}^2>0$
and $\lambda_{\rm sh}>0$, so there is no spontaneous symmetry breaking
in the singlet sector. This guarantees that the singlet scalar is
stable in the vacuum. Moreover, in the cases we will consider, the
portal coupling $\lambda_{\rm sh}$ is assumed to be very weak. This
guarantees that for light enough singlets there are no constraints
from the invisible decay width of the Higgs at the LHC.

Usually the exact value of the self-interaction $\lambda_{\rm s}$ is
considered to be irrelevant for dark matter abundance. However, as
we shall see it is of uttermost importance in determining the
initial conditions for low energy phenomena and, consequently, for
the calculation of the total dark matter yield via the freeze-in
mechanism relevant in the limit of weak portal coupling
$\lambda_{\rm sh}\lesssim 10^{-7}$.

To set the notation, we start with a brief review of the well-known freeze-in results  \cite{McDonald:2001vt, Hall:2009bx, Yaguna:2011qn,
Merle:2013wta, Klasen:2013ypa, Blennow:2013jba,
Adulpravitchai:2014xna, Merle:2014xpa,Elahi:2014fsa, Merle:2015oja, Kang:2015aqa} obtained neglecting the impacts of primordial scalar condensates generated by inflation.

In the freeze-in setup the singlet dark matter is produced from the
thermal bath of SM particles through out-of-equilibrium decays and
scatterings. For concreteness, consider the case where the dominant
process is Higgs decay into two singlet particles \cite{Hall:2009bx}
which is possible below $T_{\rm EW}$ whenever $m_{\rm h} > 2m_{\rm s}$. For a discussion of other Higgs mediated processes which for $m_{\rm h} < 2 m_{\rm s}$ could also be important, see e.g. \cite{Yaguna:2011qn}.

The evolution of number density of the singlet scalar is determined
by the Boltzmann equation \bea
\label{Boltzmann}
\dot{n}_{\rm s} + 3Hn_{\rm s} &=& \int d\Pi_h d\Pi_{\rm s_1} d\Pi_{\rm s_2}(2\pi)^4\delta^4(p_{\rm h} - p_{\rm s_1}-p_{\rm s_2}) \nn \\
&\times & \left(|\mathcal{M}|^2_{h \rightarrow ss}f_{\rm h}(1+f_{\rm s})(1+f_{\rm s}) - |\mathcal{M}|^2_{ss\rightarrow h}f_{\rm s}f_{\rm s}(1+f_{\rm h}) \right),
\eea
where $d\Pi_{\rm i} = d^3k_{\rm i}/((2\pi)^3 2E_{\rm i})$,
$\mathcal{M}$ is the transition amplitude and $f_{\rm i}$ is the usual phase space density of particle $i$. The Higgs particles are assumed to be in thermal equilibrium, and in the usual approximation one assumes that Maxwell-Boltzmann statistics can be used instead of Bose-Einstein, $f_{\rm h}\simeq e^{-E_{\rm h}/T}$.

Setting $f_s=0$ on the right hand side of Eq. (\ref{Boltzmann}) the
singlet abundance, produced at low temperatures by thermal Higgs
particles only, then becomes \cite{Hall:2009bx}
  \be \Omega_{\rm s}h^2 \approx 1.73\times
10^{27} \frac{m_{\rm s}\Gamma_{h\rightarrow ss}}{m_{\rm h}^2} =
1.73\times 10^{27} \frac{m_{\rm s}}{m_{\rm h}^2}
\left(\frac{\lambda^2_{\rm sh}\nu^2}{32\pi m_{\rm h}}\sqrt{1-4m_{\rm
s}^2/m_{\rm h}^2}\right) \ .
  \ee
In the limit, $m_{\rm s}\ll m_{\rm h}$, this yields a parametric estimate
for the coupling sufficient to produce a sizeable dark matter
abundance
  \be
\label{lowTestimate}
\lambda_{\rm sh}\simeq 10^{-11}
\left(\frac{\Omega_{\rm s}h^2}{0.12}\right)^{1/2}\left(\frac{\rm{GeV}}{m_{\rm s}}\right)^{1/2}\
.
  \ee
The implied small coupling values are compatible with the key
assumption of the freeze-in scenario that the dark matter candidate
does not thermalize with the SM background above the EW scale.

The essential approximation in the above analysis, and commonly made
in all freeze-in computations, is that the
dark matter abundance is initially negligible, i.e. $f_s=0$ on the
right hand side of Eq. (\ref{Boltzmann}) \cite{Hall:2009bx,
Yaguna:2011qn, Blennow:2013jba}. However, this may not be the
generic outcome in realistic setups where ramifications of the
inflationary stage are consistently accounted for. Indeed, scalar
dark matter candidates could easily get displaced from vacuum during
inflation and their subsequent relaxation towards the low energy
vacuum can significantly alter the freeze-in picture as we will now
turn to discuss.

\section{Primordial condensates
}
\label{sec:condensates}

\subsection{Inflationary fluctuations and their evolution}

During inflation any energetically subdominant light scalar acquires
superhorizon fluctuations proportional to the inflationary scale
$H_*$. The spectators get locally displaced from their vacuum state
and after the end of inflation the observable patch generically
features primordial spectator condensates.

Within the extended SM (\ref{scalarpot}) we are considering here,
both the Higgs and the singlet are light spectators during
inflation, $V''\ll H^2$, and get displaced from vacuum
\cite{Enqvist:2013kaa,Enqvist:2014zqa}. The typical amplitudes of
the primordial Higgs and singlet condensates are given by the root
mean square of fluctuations which for $\lambda_{\rm sh} \lesssim
\sqrt{\lambda_{\rm s} \lambda_{\rm h}}$ yields
\citep{Starobinsky:1994bd}
  \be
   \label{h,s_*}
  h_{*}={\cal O}(0.1) \frac{H_*}{\lambda_{\rm h}^{1/4}}\ ,\qquad s_{*}={\cal O}(0.1)
  \frac{H_*}{\lambda_{\rm s}^{1/4}}\ .
  \ee
The inflationary scale $H_{*}$ can be expressed in terms of the
tensor to scalar ratio $r$ and amplitude of the scalar perturbations
${\cal P}_{\zeta}$ as
 \beq
  \left(\frac{H_*}{2\pi}\right)^2=\frac{r}{8}{\cal P}_{\zeta}\ .
  \eeq
In the following we adopt these generic order of magnitude estimates
as the initial conditions for the hot big bang epoch and explore
their impacts on the singlet dark matter yield.

Here we have assumed that the inflationary scale is below the flat space instability scale $H_{*}\lesssim 10^{11}$ GeV such that $\lambda(\mu) > 0$ and the curvature induced effective Higgs mass $\xi R h^2$ is negligible. If the inflationary scale is higher, one should carefully account for the non-minimal curvature coupling which dominates the Higgs dynamics when $\lambda \rightarrow 0$ \cite{Herranen:2014cua,Espinosa:2007qp}.

Assuming an efficient reheating of the SM sector soon after the end
of inflation the Higgs condensate will be rapidly destroyed by the
thermal bath \cite{Enqvist:2014zqa} (see \cite{Enqvist:2013kaa, Enqvist:2014tta, Figueroa:2015rqa} for the decay at zero temperature and \cite{Bezrukov:2008ut, GarciaBellido:2008ab} for reheating in the context of Higgs inflation). The singlet condensate on the other hand will not feel the thermal bath for $\lambda_{\rm sh}\lesssim 10^{-7}$. Its evolution is affected by the redshifting due to expansion of space and also by out-of-equilibrium decays into singlet particles and Higgses through the portal coupling.

Neglecting the decay processes for the time being, the condensate
stays nearly constant until $H^2\sim \lambda_{\rm s} s_{*}^2$ after
which it starts to oscillate with a decreasing envelope
  \beq
  \label{s0_evolution}
   s_0(T) \simeq
   \begin{cases}
   10^{-3} \lambda_{\rm s}^{-3/8}
   r^{1/4} T\ ,
   & T \gtrsim T_{\rm trans} =   200
   \lambda_{\rm s}^{-1/8} r^{-1/4}m_{\rm s} \\
   10^{-4}\lambda_{\rm s}^{-5/16}
    r^{3/8}m_{\rm s}^{-1/2}T^{3/2}\ ,     & T\lesssim T_{\rm trans}\ .
  \end{cases}
  \eeq
For temperatures above $T_{\rm trans}$ the singlet sees an
effectively quartic potential $\lambda_{\rm s} s^4 \gg m_{\rm s}s^2$
and its energy density scales as radiation, $\rho_{\rm s}\propto
a^{-4}$. Below $T_{\rm trans}$ the quadratic mass term takes over
and the singlet energy density scales as non-relativistic matter,
$\rho_{\rm s}\propto a^{-3}$.

\subsection{The moduli problem}

If the weakly coupled singlet condensate enters the regime
$\rho_{\rm s}\sim a^{-3}$ before the matter-radiation equality its
energy density may lead to overproduction of cold dark matter. This
is essentially the well-known moduli problem encountered in a
variety of different theories for early universe physics (see e.g. \cite{Coughlan:1983ci, Banks:1995dp,Banks:1995dt}).

In the SM extension by a scalar singlet (\ref{scalarpot}) the moduli
problem constrains the viable singlet mass scale $m_{\rm s}$ from
above. Using (\ref{s0_evolution}) and requiring that the energy
density of the singlet condensate at matter-radiation transition
$T_{\rm eq} \sim 0.8$ eV does not exceed the dark matter
contribution $\Omega_{\rm s} \lesssim \Omega_{\rm DM}$ we obtain the
mass bound
  \beq
  \label{modulibound}
  \frac{m_{\rm s}}{\rm GeV} \lesssim 10^{-5}
  \left(\frac{\lambda_{\rm s}}{10^{-10}}\right)^{5/8}\left(\frac{r}{0.1}\right)^{-3/4}\
  .
  \eeq
This constraint should be regarded as an absolute upper bound as it
neglects the decay processes of condensate. The decays alleviate the
constraint by causing the condensate amplitude to decrease faster
and by depleting the singlet energy into the SM sector through Higgs
mediated processes.

\section{Freeze-in with the primordial condensates}
\label{sec:mechanism}

The primordial singlet condensate will significantly alter the
standard freeze-in picture \cite{Hall:2009bx, Yaguna:2011qn,
Blennow:2013jba}. Due to the time dependent background, singlet particles can be
produced already well above the electroweak scale where $h=0$. In this
regime the effective Boltzmann equation for the number density of
singlet particles takes the form
  \bea
\label{BoltzmannhighT}
\dot{n}_s + 3Hn_s &=& \int dPS_{\rm s_0,s,h_1,h_2}|\mathcal{M}|^2_{h \rightarrow sh} f_{\rm s_0}f_{\rm h}(1+f_{\rm s})(1+f_{\rm h})\nn \\
&+& \int  dPS_{\rm s_0,s_1,s_2,s_3} |\mathcal{M}|^2_{s \rightarrow ss}f_{\rm s_0}f_{\rm s}(1+f_{\rm s})(1+f_{\rm s})  \nn \\
&-& \int  dPS_{\rm s_0,s,h_1,h_2}|\mathcal{M}|^2_{s \rightarrow hh}f_{\rm s_0}f_{\rm s}(1+f_{\rm h})(1+f_{\rm h})  \nn \\
&+& \Gamma_{s_0\rightarrow ss}n_{\rm s_0}\ .
  \eea
Here $dPS_{\rm s_0,a,b,c}$ denotes phase space measure for the condensate $s_0$ and particles $a,b,c$ and contains the four-momentum conserving delta function. Furthermore $n_{\rm s_0}$ denotes the condensate number density,
\be
n_{\rm s_0} \equiv \int \frac{d^3k}{(2\pi)^3} f_{s_0}  \equiv \frac{ \rho_{\rm s_0}}{m_{s,{\rm eff}}}\ ,
\ee
where the singlet effective mass is defined as $m^2_{s,{\rm eff}} =V''$. As the coherently oscillating background can only decay, inverse processes where particles would go back into the condensate are not present.

The amplitudes $M_{a\rightarrow bc}$ in
\eqref{BoltzmannhighT} correspond to decay processes induced
by the oscillating condensate and $\Gamma_{s_0\rightarrow ss}
n_{s_0}$ denotes the rate for particle production directly from the
time dependent effective potential.  We approximate the amplitudes
by \cite{Abbott:1982hn, Weinberg, Ichikawa:2008ne}
  \be
  \label{Gamma1to2}
  2\pi \delta^4(p_2 -
  p_1)\mathcal{M}_{1\rightarrow 2}= \int_{-\infty}^{\infty}dt
  \langle 2 | \hat{V}(t) | 1\rangle\  ,
  \ee
where $\hat{V}(t)$ denotes the interaction Hamiltonian induced by
the singlet condensate\footnote{In the quartic regime, $\lambda_{\rm s} s^4 \gg m_{\rm s}s^2$, the system is conformal and the amplitude coincides with the Minkowski result. In the quadratic regime we keep using the Minkowski metric neglecting the small $\mathcal{O}(H/m_{\rm s})$ curvature corrections during one oscillation cycle.}. For example, in the case of $s\rightarrow
h+h$ we have
  \be
  \hat{V}(t) = -\lambda_{\rm sh}s_0(t)\int
  d^3x\hat{s}\hat{h}\hat{h}\ .
  \ee

To extract the leading contribution for the particle production
induced by the condensate, we linearize the Boltzmann equation in $f_{\rm h}$ and $f_{\rm s}$. After this the phase space integrals
can be performed and \eqref{BoltzmannhighT} reduces to
  \beq
  \label{reducedBoltzmannhighT}
  \dot{n}_{\rm s} + 3Hn_{\rm s} \simeq
   \frac{K_1\left(\frac{m_{\rm h}}{T}\right)}{K_2\left(\frac{m_{\rm h}}{T}\right)}\Gamma_{h \rightarrow sh} n_{\rm h} + \Gamma_{s_0\rightarrow ss}n_{\rm
   s_0}\ ,
  \eeq
where $K_n$ is the $n$th modified Bessel function of the 2nd kind. On the right hand side we have neglected the source terms related to the processes $s\rightarrow ss$ and $s\rightarrow hh$. These are suppressed by the small occupation numbers of singlet particles, $f_{\rm s}\ll f_{\rm h},f_{\rm s_0}$, and the latter are also kinematically heavily suppressed in the quartic regime.

The Boltzmann equation for particles (\ref{BoltzmannhighT}) is
accompanied by the corresponding equation of motion for the singlet
condensate
  \be
  \label{condensate_Boltzmann}
  \dot{n}_{\rm s_0} + 3Hn_{\rm s_0} = -
  \left(\Gamma_{s_0\rightarrow ss}+\Gamma_{s_0\rightarrow hh}\right)n_{\rm s_0} - \frac{K_1\left(\frac{m_{\rm h}}{T}\right)}{K_2\left(\frac{m_{\rm h}}{T}\right)}\Gamma_{h \rightarrow sh} n_{\rm h} \ .
  \ee
The three contributions on the right hand side correspond to energy
loss due to production of singlet particles and Higgses out of the
oscillating condensate. The decay processes have negligible effects
on the condensate motion until $\Gamma\sim H$. Up to this
point the background dynamics is therefore well described by the
solution (\ref{s0_evolution}). As $\Gamma\sim H$ the amplitude
of the condensate starts to decrease exponentially and to sufficient
accuracy we can model the process as an instant decay at
$\Gamma= H$. Consequently, the source terms in
\eqref{reducedBoltzmannhighT} vanish and the comoving singlet number
density $a^3 n_{\rm s}$ freezes to a constant. At the electroweak
transition the generation of Higgs vacuum expectation value (vev) induces an additional
contribution to the singlet number through the standard freeze-in
mechanism.

\subsection{Processes in the quartic regime}

The decay rates in the Boltzmann equation take different values
depending on whether the singlet is oscillating in the quartic
$s_0\gtrsim m_{\rm s}/\sqrt{\lambda_{\rm s}}$ or quadratic $s_0\lesssim
m_{\rm s}/\sqrt{\lambda_{\rm s}}$ regime of its potential. In terms of the
temperature, these regimes correspond respectively to $T\gtrsim
T_{\rm trans}$ and $T\lesssim T_{\rm trans}$, where the transition
temperature is given by Eq. (\ref{s0_evolution}).

In the quartic regime $T\gtrsim T_{\rm trans}$, the amplitude of the
oscillating condensate scales as $s_0\propto T$. The rate for the
$h\rightarrow sh$ process in Eqs. (\ref{reducedBoltzmannhighT}) and   (\ref{condensate_Boltzmann})
computed in the time-dependent background is given by
\be
\label{decayRates}
\Gamma^{(4)}_{h\rightarrow sh} = \frac{\lambda^2_{\rm sh}}{24\pi}\sum_{n=1}^{\infty}\frac{|s_n |^2}{m_{\rm h}}\frac{k_0}{E^{\rm h}_{k_0}+E^{\rm s}_{k_0}} \simeq 10^{-4}\lambda_{\rm
  s}^{1/2}\lambda_{\rm sh}^{2}\frac{s_0^3}{T^2}\ .
\ee
Here $k_0$ is the final state momentum satisfying the energy conservation condition $E^{\rm h}_{k_0} + E^{\rm s}_{k_0} = m_{\rm h} + n\omega$, where $\omega \simeq 0.489 m_{\rm s,eff}$ is the oscillation frequency of the singlet condensate \cite{Greene:1997fu}, and 
\be
s_0(t) =\sum_{n=-\infty}^{\infty}s_n e^{+i\omega nt} .
\ee

The other processes  in Eqs. (\ref{reducedBoltzmannhighT}) and (\ref{condensate_Boltzmann}) correspond to transitions from vacuum to two singlet or two Higgs states,
induced respectively by the interactions $\lambda_{\rm s} s_0^2(t)
s^2$ and $\lambda_{\rm sh} s_0^2(t) h^2$. The  corresponding rates are given by
  \bea
  \label{0toss}
  \Gamma^{(4)}_{s_0\rightarrow ss} &=& \frac{9\lambda^2_{\rm s}}{16\pi}\frac{m_{\rm s,eff}}{\rho_{\rm s}}\sum_{n=1}^{\infty}|\zeta_n |^2\sqrt{1-\left(\frac{2 m_{\rm \delta s}}{n\phi}\right)^2} \simeq 4\times10^{-4}\lambda_{\rm
  s}^{3/2}s_0\ \\
\label{0tohh}
 \Gamma^{(4)}_{s_0\rightarrow hh} &=&\frac{\lambda_{\rm
  sh}^{2}}{16\pi}\frac{m_{\rm s,eff}}{\rho_{\rm s}}\sum_{n=1}^{\infty}|\zeta_n|^2 \sqrt{1-\left(\frac{2 m_{\rm
  h}}{n\phi}\right)^2}\ \simeq 0,
  \eea
where $m_{\rm \delta s}^2\equiv 3\lambda_{\rm s}\langle s_0^2\rangle$, and
\be
s_0^2(t) - \langle s_0^2 \rangle = \sum_{n=-\infty}^{\infty}\zeta_n e^{-i\phi nt} ,
\ee  
with $\phi$ being the oscillation frequency of $\zeta(t)$. The rate (\ref{0tohh}) is negligible due to kinematical suppression by the thermal Higgs mass $m_{\rm h}(T) \gg m_{\rm s,eff}$.

\subsection{Processes in the quadratic regime}

In the quadratic regime the production of singlet particles directly
from the time dependent background is energetically forbidden and
consequently  $\Gamma^{(2)}_{s_0\rightarrow ss}=0$. The transition
$h\rightarrow sh$ induced by the
time-dependent condensate is also kinematically blocked in this
regime. The only possible process in Eqs. \eqref{reducedBoltzmannhighT} and \eqref{condensate_Boltzmann} is
then the production of Higgses out of the time dependent background $s_0\rightarrow hh$. The rate for this process is given by 
\be
\label{singletDecaytoHiggs}
 \Gamma^{(2)}_{s_0\rightarrow hh} = \frac{\lambda_{\rm
  sh}^{2}}{64\pi}\frac{s_0^2}{m_{\rm s}}\sqrt{1-\left(\frac{m_{\rm
  h}}{m_{\rm s}}\right)^2}\  .
\ee 
We reiterate that we are considering transitions in the time
dependent background which amounts to the kinematical condition
different for example from the standard $1\rightarrow 2$ decay in
vacuum.

From Eqs. (\ref{s0_evolution}) and (\ref{singletDecaytoHiggs}) we
obtain in the radiation dominated background the result
  \beq
  \frac{\Gamma_{s_0\rightarrow hh}^{(2)}}{H} =  \frac{\Gamma_{s_0\rightarrow hh}^{(2)}}{H} \bigg |_{t_{\rm trans}}\rule{0pt}{3ex}\bigg. \left(\frac{a_{\rm trans}}{a}\right)\  ,
   \eeq
where $t_{\rm trans}$ denotes the time of transition from quartic to
quadratic oscillations. Therefore, we immediately see that an
eventual decay of the condensate needs to take place before the
onset of quadratic oscillations. If the condensate has not decayed
by $t_{\rm trans}$ the subsequent decay rates remain  negligible and
the condensate survives undecayed.

\section{Estimating the total dark matter yield}
\label{totalYield}

The primordial singlet condensate crucially alters the standard freeze-in estimates for the abundance and properties of singlet dark matter. Depending on the strength of singlet self-coupling, the condensate may either completely decay into singlet particles or survive comprising a coherently oscillating dark matter component. The two cases could have different ramifications on structure formation, see e.g. \cite{Hu:1998kj,Hu:2000ke,Amendola:2005ad,Marsh:2013ywa}. The fate of the singlet condensate also
affects the eventual decay channels of the singlet provided that the
portal sector contains additional degrees of freedom such as
fermions with Yukawa couplings ${\cal L} = g s {\bar
\psi}\psi $.  

In addition to the primordial condensates, singlet particles are produced at the electroweak transition through Higgs decays $\lambda_{\rm sh}\nu h ss$. However, as opposed to the standard freeze-in estimates the singlet occupation numbers need not be small which could significantly affect the process. We leave a detailed analysis on this question to future work. In what follows we concentrate only on the abundace of singlet dark matter generated through the primordial condensates. 

We also note that the observational bounds are significantly different depending on whether the singlet constitutes isocurvature or adiabatic dark matter. While the component sourced by the primordial condensates clearly is isocurvature, the situation is less clear when the production of singlet particles through Higgs decay is important. Any additional couplings between the SM and the portal sector would also affect the situation. We will not dwell further on this important issue but simply choose to present both the adiabatic and isocurvature bounds in what follows.

\subsection{Case I: The condensate decays completely}
\label{case1}

As we have seen
above, the condensate can decay only if the decay takes place in the
quartic regime where $\lambda_{\rm s} s_0^2 \gg m^2_{\rm s}$. With the rates given
in Eqs. (\ref{decayRates}) , (\ref{0toss}) and (\ref{0tohh}) the Boltzmann
equations for the particles (\ref{reducedBoltzmannhighT}) and
condensate (\ref{condensate_Boltzmann}) in this regime are given by
  \baq
  \label{caseI_particles}
  \dot{n}_{\rm s} + 3Hn_{\rm s} &=&
   \frac{K_1\left(\frac{m_{\rm h}}{T}\right)}{K_2\left(\frac{m_{\rm h}}{T}\right)}\Gamma_{h \rightarrow sh} n_{\rm h} + \Gamma_{s_0\rightarrow ss}n_{\rm
   s_0}\ ,\\
 \label{caseI_condensate}
    \dot{n}_{\rm s_0} + 3H n_{\rm s_0} &=&  - \frac{K_1\left(\frac{m_{\rm h}}{T}\right)}{K_2\left(\frac{m_{\rm h}}{T}\right)}\Gamma_{h \rightarrow sh} n_{\rm h} -  
  \Gamma_{s_0\rightarrow ss} n_{\rm s_0} \ .
  \eaq

Using that $n_{\rm s_0} = \rho_{\rm s}/ m_{\rm s,{\rm eff}} = \lambda_{s} s_0^3 /(4\sqrt{3})$ we can express the $s_0$ dependence of  $\Gamma_{h\rightarrow sh}$ (\ref{decayRates})  in terms of $n_{\rm s_0}$ and rewrite the corresponding term in the condensate Boltzmann equation (\ref{caseI_condensate}) as 
  \beq
\label{tildegamma}
  \frac{K_1\left(\frac{m_{\rm h}}{T}\right)}{K_2\left(\frac{m_{\rm h}}{T}\right)}\Gamma_{h \rightarrow sh} n_{\rm h} \simeq  10^{-4} \frac{K_1\left(\frac{m_{\rm h}}{T}\right)}{K_2\left(\frac{m_{\rm h}}{T}\right)}  4\sqrt{3}\lambda_{\rm sh}^2\frac{n_{\rm h}}{T^2} n_{\rm s_0} \equiv \tilde{\Gamma}_{h\rightarrow sh} n_{\rm s_0}\ .
  \eeq
The solution for equation of motion (\ref{caseI_condensate}) can then be written
in the implicit form
 \be
 \label{rhos_implicit}
 n_{\rm s_0} = n_{\rm s_0, {\rm osc}}\left(\frac{a_{\rm osc}}{a}\right)^3
   \exp\left(-\int_{a_{\rm osc}}^a
  \frac{da}{a}\left(\frac{\Gamma_{s_0\rightarrow ss}(s_0)}{H}+\frac{\tilde{\Gamma}_{h\rightarrow sh}}{H} \right)\right)
  \  ,
  \ee
where the quantities with subscript osc are evaluated at $t_{\rm osc}$ denoting the onset of quartic oscillations,
$H_{\rm osc} \sim \lambda_{\rm s} s^2$ and  only the first term inside the integral depends on $n_{\rm s_0}$.

When $\Gamma_{s_0\rightarrow ss}, \tilde{\Gamma}_{h\rightarrow sh}\lesssim H$ the particle
production clearly has a negligible effect on the motion of the condensate.
Using that\footnote{See Eqs.  (\ref{0toss}),  (\ref{tildegamma}) and (\ref{s0_evolution}), respectively, for the following scalings.}  $\Gamma_{s_0\rightarrow ss}\propto s_0 $, and $\tilde{\Gamma}_{h\rightarrow sh}\propto T$
and substituting $s_0\propto a^{-1}$
under the integral in Eq. (\ref{rhos_implicit})  the integral can then be performed. This yields the approximative 
explicit solution
  \beq
  n_{\rm s_0} \simeq 10^{-4}
  \frac{H^3_{*}}{\lambda_{\rm s}^{1/4}}\left(\frac{a_{\rm osc}}{a}\right)^3{\rm exp}\left(-\frac{\Gamma_{s_0\rightarrow ss}(s_0)}{H}-\frac{\tilde{\Gamma}_{h\rightarrow sh}}{H}\right)\ .
  \eeq
Here $H_{*}$ denotes the inflationary Hubble rate at the horizon
exit of observable modes.

\begin{figure}[htb]
\begin{center}
\includegraphics[width=.6\textwidth]{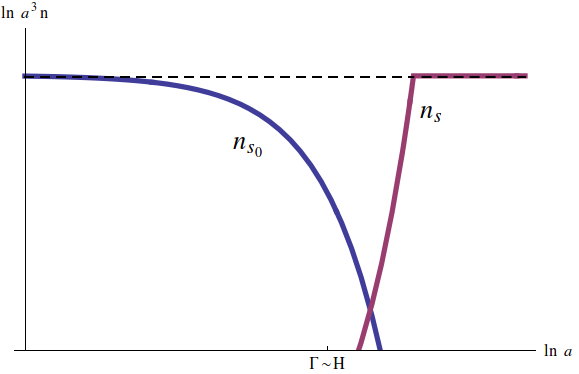}
\caption{A schematic representation of the evolution of singlet number
density and how it is divided between the condensate and
excitations. The dashed line shows the constant comoving number density, the purple curve shows the
number density of singlet particles and the blue one shows the number
density of the condensate. The figure corresponds to the case where
the interactions are sufficiently strong to allow for the condensate
to decay away ($\Gamma\sim H$) before the quadratic part of the
singlet potential starts to dominate the evolution (at
$T_{\rm{trans}}$). If the condensate survives until
$T_{\rm{trans}}$, it practically does not decay at all and its comoving number density remains almost constant.}
\label{DMscaling}
\end{center}
\end{figure}

For the weak portal couplings $\lambda_{\rm sh}\lesssim 10^{-7}$ we are considering here, we generically have $\Gamma_{s_0\rightarrow ss} \gg  \tilde{\Gamma}_{h\rightarrow sh}$ if the condensate is to decay in the quartic regime. The decay time is then dictated by  $\Gamma_{s_0\rightarrow ss}$ and the condensate decays practically
instantaneously at $H_{\rm dec}=\Gamma_{s_0\rightarrow ss}$
depleting all its energy into singlet particles. The solution of
Boltzmann equation for singlet particles then reads 
  \beq
  \label{ns_caseI}
  n_{s} \simeq \frac{1}{a^{3}}\left(\frac{K_1\left(\frac{m_{\rm h}}{T}\right)}{K_2\left(\frac{m_{\rm h}}{T}\right)}\frac{\Gamma_{h\rightarrow
  sh}}{\Gamma_{s_0\rightarrow ss}} n_{\rm h} a^{3}+ n_{\rm s_0} a^{3}\right)_{t=t_{\rm
  osc}} \ ,
  \eeq
showing that the comoving number density freezes to a constant value
as the time-dependent background vanishes. The behaviour is
illustrated in Figure \ref{DMscaling}. 

The distribution of the generated singlet particles is peaked at $k_{*}/a_{\rm dec} \sim \sqrt{3\lambda_{\rm s}} s_0$. As $\sqrt{3\lambda_{\rm s}} s_0 > m_{\rm s}$, after the decay  of the condensate the produced singlet particles constitute effectively relativistic matter until $k_*/a \sim m_{\rm s}$ and they become non-relativistic. If the decay of the condensate is complete, the corresponding transition temperature $T_{\rm trans}$ is given by Eq. (\ref{s0_evolution}) as before. For $T<T_{\rm trans}$ the energy density of singlet particles is thus given by  $\rho_{\rm s} = m_{\rm s} n_{\rm s}$ which yields
   \be
\label{condensateDecayAbundance}
\left(\frac{\Omega_{\rm s} h^2}{0.12}\right) = 10^4\lambda_{\rm
s}^{-7/4}\lambda_{\rm
sh}^{2}\left(\frac{r}{0.1}\right)^{1/2}\left(\frac{m_{\rm s}}{\rm GeV}\right) + 10\lambda_{\rm
s}^{-5/8}\left(\frac{r}{0.1}\right)^{3/4}\left(\frac{m_{\rm s}}{\rm GeV}\right).
  \ee
for the present abundance of singlet dark matter.

The parametric dependence of the dark matter yield is illustrated in
Figures \ref{highTyield} and \ref{highTyield2}. The red domain in the
figures marks the regime where the condensate decays completely and
dark matter consists of singlet particles. This domain exists only
for sufficiently large values of the singlet self-coupling
$\lambda_{\rm s}$ or the portal coupling $\lambda_{\rm sh}$. For smaller
couplings the condensate decay never becomes efficient and the final
dark matter component will consist of the singlet condensate instead
of particles.

As reviewed in Sec. \ref{sec:model}, for superweak portal coupling the correct
dark matter abundance can be generated at low temperature \cite{McDonald:2001vt}. 
Here we have demonstrated that the primordial singlet condensate crucially affects the abundance and
properties of singlet dark matter providing non-trivial boundary condition for the low temperature dark matter production. 
We leave for future work the careful matching of the low temperature particle production with the high temperature freeze-in we have described here.

\subsection{Case II: The condensate survives}
\label{case2}

After the transition from quartic to quadratic oscillations the
decay rate of the singlet condensate decreases faster than the
Hubble rate. In the parameter range where $\Gamma_{s_0\rightarrow ss} \ll H$ until the end of the quartic epoch the condensate therefore practically does not decay at all but is merely redshifted according to Eq. (\ref{s0_evolution}).

In more detail, the corresponding Boltzmann equation for singlet particles
\eqref{reducedBoltzmannhighT} in the quartic regime $T\gtrsim
T_{\rm trans}$ is given by,
  \be
  \label{highTeq}
  \frac{dY^{(4)}_{\rm s}}{dT} = -\frac{K_1\left(\frac{m_{\rm h}}{T}\right)}{K_2\left(\frac{m_{\rm h}}{T}\right)}\frac{\Gamma^{(4)}_{h\rightarrow sh}}{Hs_{\rm b}T}n_{\rm h}-\frac{\Gamma^{(4)}_{s_0\rightarrow ss}}{Hs_{\rm b}T}n_{\rm s_0}\  ,
  \ee
where $Y_{\rm s}\equiv n_{\rm s}/s_{\rm b}$ denotes the singlet
number density normalized by the entropy density of the bath $s_{\rm
b}$ and where we used $\dot{T}\simeq-HT$, which is an excellent approximation above the EW scale. 
With the rates given in Eqs. (\ref{decayRates}),  (\ref{0toss}) and (\ref{0tohh}), the solution of Eq. (\ref{highTeq}) is
 \be
 \label{DMsolution}
  Y^{(4)}_{\rm s}(T) = \left(4\times 10^4\lambda_{\rm sh}^2\lambda_{\rm s}^{-5/8}\left(\frac{r}{0.1}\right)^{3/4} +5\times 10^2\lambda_{\rm s}^{1/2}\left(\frac{r}{0.1}\right) \right)\frac{\rm GeV}{T} .
 \ee
 
After the transition to the quadratic regime $T\lesssim T_{\rm
trans}$ the kinematical suppression renders the singlet particle production negligible and the comoving particle number freezes to a constant value. The corresponding present particle abundance is
  \bea
  \label{finalabundancequadr}
  \left(\frac{\Omega_{\rm s} h^2}{0.12}\right) &=&
  2.286\times10^9\left(\frac{m_{\rm s}}{\rm{GeV}}\right) Y^{(4)}_{\rm
  s}(T_0)\ .
  \eea

\begin{figure}[htb]
\begin{minipage}[b]{0.45\linewidth}
\centering
\includegraphics[width=\textwidth]{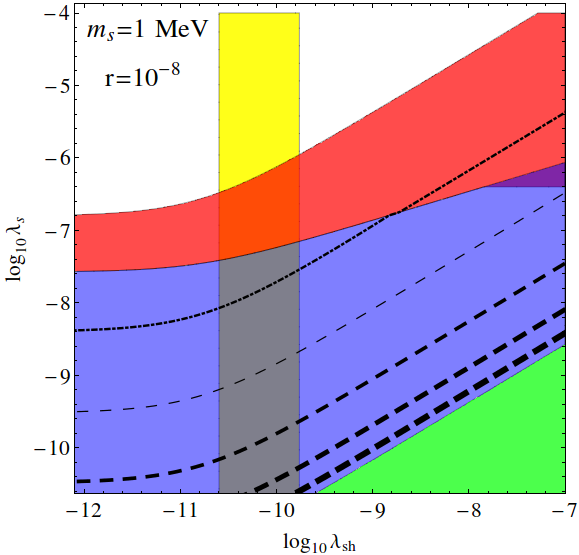}
\end{minipage}
\begin{minipage}[b]{0.45\linewidth}
\centering
\includegraphics[width=\textwidth]{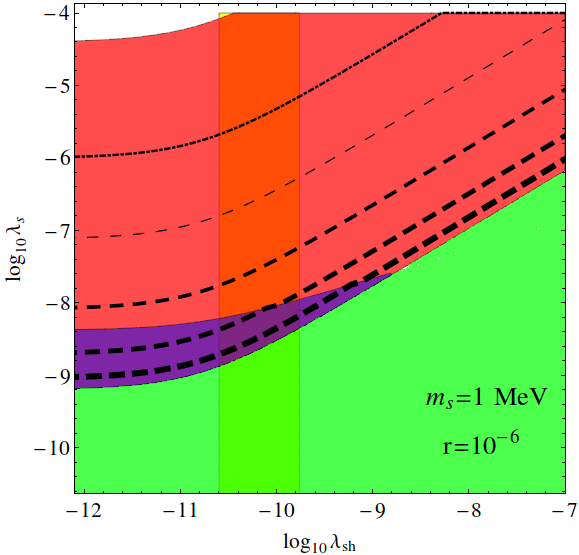}
\end{minipage}
\begin{minipage}[b]{0.45\linewidth}
\centering
\includegraphics[width=\textwidth]{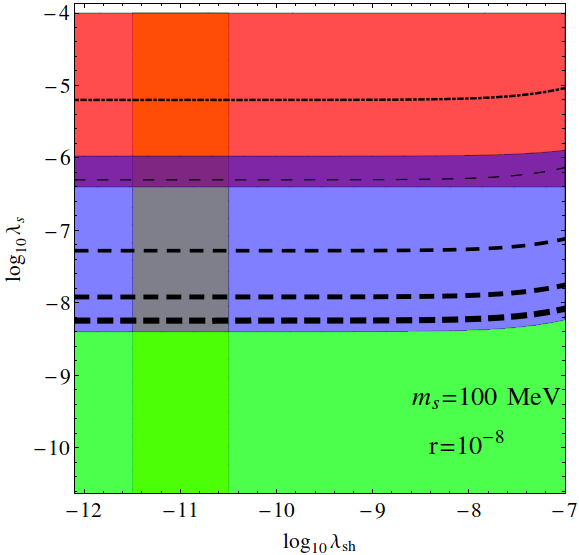}
\end{minipage}
\begin{minipage}[b]{0.45\linewidth}
\centering
\includegraphics[width=\textwidth]{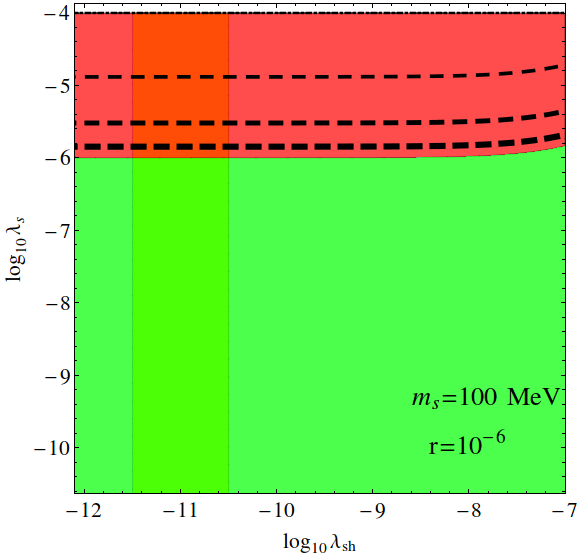}
\end{minipage}
\centering
\caption{The final dark matter yield, shown in terms of the portal coupling $\lambda_{\rm sh}$ and the self-interaction coupling $\lambda_{\rm s}$.
Different regions where the model components (singlet condensate,
singlet particles) produce 0.1-100\% of today's dark matter abundance
are shown in red (particles only), blue (condensate only) and purple
(both particles and condensate); see the main text for further
explanations. The slashed contours refer to 80\%, 50\%, 20\% and 5\%
abundances, listed from thickest to thinnest contour. The green region shows where the singlet would constitute more than 100\% of today's DM abundance. The dotted line denotes the upper bound ($\sim$1\%) for an isocurvature DM component. The vertical yellow band shows the standard scenario where only the low temperature processes produce 1-100\% of today's DM abundance.}
\label{highTyield}
\end{figure}

\begin{figure}[htb]
\begin{minipage}[b]{0.45\linewidth}
\centering
\includegraphics[width=\textwidth]{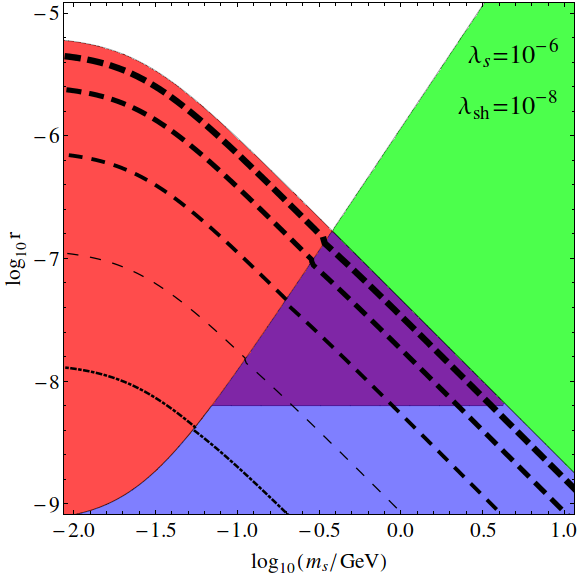}
\end{minipage}
\begin{minipage}[b]{0.45\linewidth}
\centering
\includegraphics[width=\textwidth]{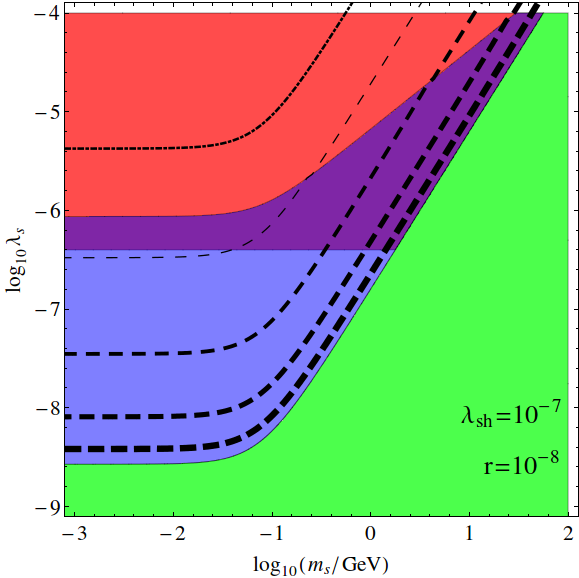}
\end{minipage}
\centering \caption{The final dark matter yield, shown in terms of
the singlet mass $m_{\rm s}$ and tensor-to-scalar ratio $r$ (left
panel) or the self-interaction coupling $\lambda_{\rm s}$ (right
panel). The different colours and contours are the same as in Figure
\ref{highTyield}.}
\label{highTyield2}
\end{figure}

The final yield corresponding to the correct DM abundance today is
depicted in Figures \ref{highTyield} and \ref{highTyield2} for
representative values of model parameters. The regions below the red
domains correspond to the case where the condensate survives. The purple areas just over the blue regions depict the case II, where there is a small fraction of relic density also in the singlet particles.
The dotted line denotes the upper bound for an isocurvature DM component, allowing only $\sim$1\% of non-thermally generated DM contributing to the observed total DM abundance \cite{Ade:2015xua}.
The dashed lines, from thinnest to thickest denote 5\%, 20\%, 50\% and 80\% abundances, respectively. For example, in Figure \ref{highTyield} the upper
boundary of the red region corresponds to 0.1\% abundance and the
lower boundary of the blue band to 100\%.  The yellow vertical band
shows the standard freeze-in scenario \cite{McDonald:2001vt,
Hall:2009bx, Yaguna:2011qn}, where only the low temperature
processes ($h\rightarrow ss$) produce 1-100\% of the present DM
abundance, assuming $f_{\rm s}=0$ at $T=T_{\rm{EW}}$. Note that the portal
coupling is always required to be $\lambda_{\rm sh}\lesssim 10^{-7}$
in order to avoid singlet thermalization before the electroweak
scale.

As we have discussed above, one important consequence of the primordial singlet condensate is that it will lead to isocurvature perturbations which are heavily constrained by their imprints on the CMB anisotropies \cite{Ade:2015xua}. For the specific SM extension (\ref{scalarpot}) we have concentrated on here, this constrains the singlet dark matter component to constitute at most only a very small fraction of the total dark matter abundance,  as shown in Figures \ref{highTyield} and \ref{highTyield2}. 

However, extending beyond the simplistic scenario (\ref{scalarpot}) these constraints could be easily relaxed.  In particular, we can entertain a thought that there are additional fields and interactions in the portal sector. These should not affect the high temperature evolution we have considered here, but could provide additional channels for the singlet sector to interact with the SM fields at temperatures below the electroweak transition. These interactions could convert the isocurvature modes into the observed adiabatic perturbations. A concrete possibility would be a superheavy scalar attaining a non-zero vev only at low temperatures.

\section{Conclusions and outlook}
\label{sec:conclusions}

We have considered in detail the consequences of inflationary
initial conditions on the dynamics of (extended) scalar sectors with
light excitations. Concretely, we considered the simple benchmark model
where the dark matter is constituted by a $Z_2$ symmetric real
scalar field. 
Our main results are equations \eqref{condensateDecayAbundance} and \eqref{finalabundancequadr} together with Figures \ref{highTyield} and \ref{highTyield2}. They have strong implications on presently
popular models of dark matter production via the freeze-in
mechanism.

One particularly important feature we have uncovered is that, contrary to what is often assumed, $\lambda_{\rm s}$ is not irrelevant for the production of the dark matter abundance. 
Another important feature is that significant particle production via the freeze-in mechanism is possible at high-temperatures, above the electroweak scale, due to primordial condensates following from the initial conditions set by inflation. Consequently, the initial conditions for the singlet abundance to be negligible at $T\sim T_{\rm EW}$ may not be valid. More generally, our analysis has revealed a novel connection how fundamental inflationary physics is imprinted on the dark matter abundance.\footnote{For a related effect, see \cite{Dev:2014tla}.}

Within the singlet scalar extension of SM we have shown how the freeze-in scenario severely constrains
both the self-interaction coupling $\lambda_{\rm s}$ and portal coupling $\lambda_{\rm sh}$. Our results can be also
extended to provide new constraints on models in which the scalar acts only as a mediator and decays further to the actual DM particle such as a sterile neutrino. We will consider scenarios beyond the simple benchmark one treated in this paper in a forthcoming work.

\section*{Acknowledgements}
We thank K. Kainulainen for discussions. This work was financially supported by the Academy of Finland projects 257532 (SN) and 267842 (KT). TT is supported by the Research Foundation of the University of Helsinki.

\bibliography{freezein}

\end{document}